\documentclass[onecolumn,floatfix,superscriptaddress,a4paper,
               showpacs,showkeys,nofootinbib,preprint]{revtex4-1}

\usepackage{epsfig}
\usepackage{latexsym}
\usepackage{xspace}
\usepackage{hyperref}
\usepackage[latin2]{inputenc}
\usepackage{indentfirst}
\usepackage{enumerate}

\usepackage{amsmath}
\usepackage{amssymb}
\usepackage[english]{babel}
\usepackage{url}

\newcommand{\eq}[1]{\begin{align} #1 \end{align}}

\begin{document}

\title{
Beth-Uhlenbeck approach for repulsive interactions
between baryons
in a hadron gas
}
\author{Volodymyr Vovchenko}
\affiliation{
Institut f\"ur Theoretische Physik,
Goethe Universit\"at, D-60438 Frankfurt am Main, Germany}
\affiliation{
Frankfurt Institute for Advanced Studies, Giersch Science Center,
D-60438 Frankfurt am Main, Germany}
\affiliation{Department of Physics, Taras Shevchenko National University of Kiev, 03022 Kiev, Ukraine}
\author{Anton Motornenko}
\affiliation{
Institut f\"ur Theoretische Physik,
Goethe Universit\"at, D-60438 Frankfurt am Main, Germany}
\affiliation{
Frankfurt Institute for Advanced Studies, Giersch Science Center,
D-60438 Frankfurt am Main, Germany}
\author{Mark~I.~Gorenstein}
\affiliation{
Bogolyubov Institute for Theoretical Physics, 03680 Kiev, Ukraine}
\affiliation{
Frankfurt Institute for Advanced Studies, Giersch Science Center,
D-60438 Frankfurt am Main, Germany}
\author{Horst Stoecker}
\affiliation{
Institut f\"ur Theoretische Physik,
Goethe Universit\"at, D-60438 Frankfurt am Main, Germany}
\affiliation{
Frankfurt Institute for Advanced Studies, Giersch Science Center,
D-60438 Frankfurt am Main, Germany}
\affiliation{
GSI Helmholtzzentrum f\"ur Schwerionenforschung GmbH, D-64291 Darmstadt, Germany}

\date{\today}

\begin{abstract}
The quantum mechanical Beth-Uhlenbeck (BU) approach for repulsive hard-core interactions between baryons is applied to the thermodynamics of a hadron gas.
The second virial coefficient $a_2$ -- the ``excluded volume'' parameter -- calculated within the BU approach is found to be temperature dependent, and it differs dramatically from the classical excluded volume (EV) model result.
At temperatures $T =100-200$~MeV, the widely used classical EV model underestimates the EV parameter for nucleons at a given value of the nucleon hard-core radius by large factors of 3-4.
Previous studies, which employed the hard-core radii of hadrons as an input into the classical EV model, have to be re-evaluated using the appropriately rescaled EV parameters.
The BU approach is used to model the repulsive baryonic interactions in the hadron resonance gas (HRG) model.
Lattice data for the second and fourth order net baryon susceptibilities are described fairly well when the 
temperature dependent BU baryonic excluded volume parameter corresponds
to nucleon hard-core radii of $r_c = 0.25-0.3$~fm.
Role of the attractive baryonic interactions is also considered.
It is argued that HRG model with a constant baryon-baryon EV parameter $v_{NN} \simeq 1$~fm$^3$ provides a simple 
yet efficient description of baryon-baryon interaction in the crossover temperature region.
\end{abstract}



\maketitle

\section{Introduction}

The properties of QCD at high densities and temperatures are studied experimentally and theoretically using
relativistic heavy-ion collisions.
Lattice QCD simulations, hydrodynamic and transport models are among the tools employed.
Lattice QCD observables at zero chemical potential, $\mu_B = 0$, and at moderate temperatures, $T \lesssim 150$~MeV, are reasonably well described by the ideal hadron resonance gas (HRG) model~\cite{Karsch:2003vd,Karsch:2003zq,Tawfik:2004sw,Borsanyi:2011sw,Bazavov:2012jq,Bellwied:2015lba,Bellwied:2013cta}.

The standard HRG model assumes that microscopic system states consist of non-interacting hadrons and resonances~\cite{Hagedorn:1965st}. In accord with the arguments based on the S-matrix approach~\cite{Dashen:1969ep,Venugopalan:1992hy,Lo:2017sde}, this HRG model includes attractive interactions between hadrons which lead to the formation of resonances.
The resonances in HRG can also be treated within the K-matrix approach~\cite{Wigner:1946zz,Wigner:1947zz,Chung:1995dx}, in particular for the case of the overlapping resonances~\cite{Chung:1995dx,Wiranata:2013oaa}.
More realistic hadronic models take into account the presence of both, attractive and repulsive interactions between the constituent hadrons.
Repulsive interactions in the HRG model had previously been considered in the framework of the relativistic Mayer's~(cluster) and virial expansions~\cite{Venugopalan:1992hy}, via repulsive mean fields~\cite{Olive:1980dy,Olive:1982we}, and via excluded volume (EV) corrections~\cite{Hagedorn:1980kb,Gorenstein:1981fa,Hagedorn:1982qh,Kapusta:1982qd,Rischke:1991ke,Anchishkin:1991ig,Anchishkin:1995np}.
In particular, the effects of EV interactions between hadrons on HRG thermodynamics~\cite{Satarov:2009zx,Andronic:2012ut,Bhattacharyya:2013oya,Albright:2014gva,Vovchenko:2014pka,Albright:2015uua,Zalewski:2015yea,Alba:2016fku} and on observables in heavy-ion collisions~\cite{Yen:1997rv,Yen:1998pa,BraunMunzinger:1999qy,Cleymans:2005xv,Begun:2012rf,Fu:2013gga,Vovchenko:2015cbk,Vovchenko:2016ebv,Alba:2016hwx,Satarov:2016peb} had extensively been studied in the literature.
Recently, repulsive interactions have received renewed interest in the context of lattice QCD data on fluctuations of conserved charges.
It was shown that large deviations of several fluctuation observables from the \emph{ideal} HRG baseline could well be interpreted in terms of repulsive baryon-baryon interactions~\cite{Vovchenko:2016rkn,Vovchenko:2017cbu,Huovinen:2017ogf,Vovchenko:2017xad}.

The total system volume of the thermodynamic systems is substituted in the EV model by the total available volume, i.e. $V \to V - v\,N$, where $v$ is the excluded volume parameter of a single particle.
The microscopic background of the EV model corresponds to repulsive hard-core interactions.
$v$ is connected to the microscopic hard-core radius $r_c$ of a particle as $v = \displaystyle \frac{16\pi}{3} r_c^3$ in the single-component EV model~\cite{LL,Huang}.
In the context of hadronic physics applications it is, however,  often overlooked that the above relation between $v$ and $r_c$ is inherently \emph{classical}, i.e. all quantum mechanical effects on the hard-core interaction are ignored.
Such an approximation may be justified when the thermal de Broglie wavelength of the constituent particles is much smaller as compared to their hard-core radius, i.e. $\lambda_{\rm dB} \ll r_c$, which is the case for the EV model applications in classical physics.
When $\lambda_{\rm dB} \gtrsim r_c$, however, the classical approximation breaks down, as shown in Refs.~\cite{Kostyuk:2000nx,Typel:2016srf} for the case of spinless particles.
A simple estimate for nucleons ($m_N\cong 938$~MeV is assumed in this paper)
yields a de Broglie wavelength $\lambda_{\rm dB} = \sqrt{2\pi/(m_NT)} \simeq 1.3$~fm at $T = 150$~MeV. 
This value is much larger than typical nucleon hard-core radii, with values of about $r_c = 0.2-0.8$~fm often employed by practitioners of the EV model~\cite{Yen:1997rv,Vovchenko:2014pka,Yen:1998pa,BraunMunzinger:1999qy,Cleymans:2005xv,Andronic:2012ut,Begun:2012rf,Fu:2013gga,Vovchenko:2015cbk,Vovchenko:2016ebv}.
These $\lambda_{\rm dB}$ values are even larger at smaller temperatures.
Thus, the hard-core interactions between hadrons are expected to be significantly affected by quantum mechanical effects at these temperatures.

The present paper explores quantum mechanical effects on the 2nd virial coefficient in systems of baryons with hard-core interactions
in the framework of the Beth-Uhlenbeck (BU) approach~\cite{Beth:1937zz}.
The results are contrasted with the classical EV model.
The
classical EV model, as well as its virial- and cluster expansions, are elaborated in Sec.~\ref{sec:claEV}.
Sec.~\ref{sec:BUEV} describes, within the BU approach, the effects of hard-core interactions on the 2nd virial coefficient in the system of nucleons.
The applications of the BU approach to
the HRG model with repulsive baryon-baryon interactions are discussed in Sec.~\ref{sec:HRG}.

\section{Classical excluded-volume model}
\label{sec:claEV}

Short-range repulsive interactions are modeled in the classical EV model by substituting the total volume by the available volume, i.e. $V \to V - vN$, where $N$ is the total number of particles.
This substitution results in the well known van der Waals equation of state
\eq{\label{eq:ev}
p^{\rm ev}(T,n) = \frac{Tn}{1 - vn},
}
in which the attractive van der Waals interactions are omitted.
Here $n \equiv N / V$ is the particle number density.

The pressure function, $p(T,n)$, can be written in form of the \emph{virial expansion}~\cite{mayer,Huang,greiner}
\eq{\label{eq:virialexp}
p(T,n) = T \, \sum_{k = 1}^{\infty} \, a_k(T) \, n^k.
}
Here $a_k$ are the virial coefficients.
These virial coefficients are temperature independent in the classical EV model, as follows from Eq.~\eqref{eq:ev}:
\eq{
a_{k}^{\rm ev} = v^{k-1}.
}

The 2nd virial coefficient can be related to the hard-core radius $r_c$ of a given constituent.
If quantum mechanical effects are neglected, $a_2(T)$ is related to the 2-body interaction potential by
\eq{\label{eq:b2}
a_2(T) = \frac{1}{2} \int d^3 r \, \left\{1 - \exp\left[- \frac{U(r)}{T} \right] \right\}.
}
A repulsive hard-core potential reads
\eq{\label{eq:Uhc}
U(r) =
\begin{cases}
\infty, & \quad  \textrm{if}~ r < 2 r_c\\
0, & \quad \textrm{if}~  r > 2 r_c.
\end{cases}
}
Substituting $U(r)$ from \eqref{eq:Uhc} into \eqref{eq:b2} yields
\eq{\label{eq:evcl}
a_2^{\rm ev} = v = \frac{16 \pi r_c^3}{3}.
}

Let us discuss the Mayer's \emph{cluster} expansion of the pressure in the EV model. This expansion is in terms of the powers of the fugacity, $\lambda = e^{\mu / T}$. It will be used below for the comparison with the BU approach.
The cluster expansion 
is written as~\cite{mayer,Huang,greiner}
\eq{\label{eq:clusterexp}
p(T,\mu) = T \, \sum_{k = 1}^{\infty} \, b_k(T) \, [g \, \phi(T;m) \, \lambda]^k = T \, \sum_{k = 1}^{\infty} \, b_k(T) \, z^k.
}
Here $z \equiv g \, \phi(T;m) \, \lambda$ is the absolute activity, which can be considered as the density of the ideal gas with Boltzmann statistics at a given $T$-$\mu$ pair, and $b_k(T)$ are the cluster integrals, i.e. the coefficients of the Mayer's cluster expansion in fugacities~(see, e.g., Chapter 10 in Ref.~\cite{Huang}).
Function $\phi(T;m)$ is expressed via the modified Bessel function $K_2$,
\eq{\label{phi}
\phi(T;m) =  \frac{m^2 \, T}{2\pi^2} \, K_2\left( m \over T \right),
}
where we assumed the relativistic dispersion relation $\varepsilon(k) = \sqrt{m^2+k^2}$.

The pressure of the EV model in $T$-$\mu$ variables is given in terms of the transcendental equation ${p^{\rm ev}(T,\mu) = p^{\rm id} (T, \mu - v \, p^{\rm ev})}$. Expansion of the EV model pressure around the ideal gas pressure $p^{\rm id} (T, \mu)$ yields
\eq{\label{eq:clev}
p^{\rm ev} (T, \mu) & = p^{\rm id} (T, \mu - v \, p^{\rm ev}) \nonumber \\
& = p^{\rm id} (T, \mu) - n^{\rm id} (T, \mu) \, v \, p^{\rm ev} (T, \mu) + \ldots
\nonumber \\
& = T \, g \, \phi(T;m) \, \lambda - T \, v \, [g \, \phi(T;m)]^2 \, \lambda^2 + O(\lambda^3),
}
where $g$ is the internal degeneracy factor~(for nucleons $g_N = 4$). 
Note, that the effects of quantum statistics were neglected in the final line in Eq.~(\ref{eq:clev}).
Only the behavior of the 2nd cluster or virial coefficients is analyzed in the present work, therefore, the expansion in Eq.~\eqref{eq:clev} is written only up to the 2nd order.

Comparison of Eqs.~\eqref{eq:clev} and \eqref{eq:clusterexp} yields
\eq{\label{eq:b2ev}
b_2^{\rm ev} = -v = -a_2^{\rm ev}.
}
Thus, the 2nd cluster integral is straightforwardly connected to the excluded volume parameter $v$.

\section{Beth-Uhlenbeck approach}
\label{sec:BUEV}

\subsection{Formalism}

Both the virial~\eqref{eq:virialexp} and the cluster~\eqref{eq:clusterexp} expansion can be applied to describe interactions in a quantum system.
If particles interact elastically and do not form bound states, the 2nd cluster integral is given by the generalized BU formula~\cite{Dashen:1969ep,Venugopalan:1992hy,Kostyuk:2000nx}\footnote{The ideal quantum gas contribution to $b_2(T)$, found to be negligible for the applications considered in the present paper, is neglected for simplicity.}
\eq{\label{eq:BUgen}
b_2(T) = [g \, \phi(T;m)]^{-2} \, \frac{T}{2\pi^3} \, \int_{2m}^{\infty} d\varepsilon \, \varepsilon^2 \, \operatorname{K}_2(\varepsilon/T) \, \sum_{Q} \, g_Q \, \frac{d \delta_Q (\varepsilon)}{d \varepsilon}.
}
Here the integral runs over all values 
of the invariant mass $\varepsilon$ of two particles in the center-of-mass frame.
The sum in the integrand is taken over all relevant channels of all two-particle states, which are characterized by a set of quantum numbers $Q$. The specific definition of $Q$ depends on a particular system studied~(see below).
$\delta_Q (\varepsilon)$ is the corresponding scattering phase shift for channel $Q$.
Equation~\eqref{eq:BUgen} assumes relativistic dispersion relation $\varepsilon(k) = \sqrt{m^2+k^2}$ between energy and momentum.

Let us consider a system of interacting nucleons.
For nucleon-nucleon scattering, the corresponding set of quantum numbers is $Q = (T,S,L,J)$: isospin $T = 0,1$; spin $S = 0,1$; orbital momentum $L$; total angular momentum $J$, which takes the values ${|L-S| < J < (L+S)}$.
The value of the orbital momentum $L$ determines the symmetry of the coordinate part of the two-nucleon wave function with respect to the exchange of the coordinates of two nucleons: For even values of $L$, it is symmetric, while for odd values of $L$, it is antisymmetric.
The total two-nucleon wave function is antisymmetric with respect to the exchange of their indices. Thus, $L$ takes odd values if the spin-isospin part is symmetric, and even values otherwise.
The sum in \eqref{eq:BUgen} goes over all possible $(T,S,L,J)$  values that are consistent with the above restrictions.

The scattering phase shifts are well known for the hard-sphere scattering potential~\eqref{eq:Uhc}.
They depend on the orbital angular momentum $L$ and are given by~\cite{Schiff}
\eq{\label{eq:deltahc}
\delta_L^{\rm hc} (\varepsilon) = \arctan\left[ \frac{j_L(2 r_c\,q)}{n_L(2 r_c\,q)} \right].
}
Here $q \equiv q(\varepsilon)$ is the momentum of a constituent particle in the c.m. frame, $j_L$ and $n_L$ are spherical Bessel functions.
Relativistic dispersion relation is employed in the present work, therefore $q(\varepsilon) = \frac{1}{2} \sqrt{\varepsilon^2 - (2m_N)^2}$.
Thus, the expression for the 2nd cluster integral for the nucleon system with a hard-core interaction can be written as
\eq{\label{eq:BUNN1}
b_2^{NN}(T) & = [g_N \, \phi(T;m_N)]^{-2} \, \frac{T}{2\pi^3} \, \int_{2m_N}^{\infty} d\varepsilon \, \varepsilon^2 \, \operatorname{K}_2(\varepsilon/T) \nonumber \\
& \qquad \times
\sum_{T=0,1} \, \sum_{S=0,1} \, \sum_{L} \, \sum_{J=|L-S|}^{L+S}\, (2\,T + 1) \, (2\,J+1) \, \frac{d \delta_L^{\rm hc} (\varepsilon)}{d \varepsilon}.
}
Integration by parts yields
\eq{\label{eq:BUNN2}
b_2^{NN}(T) & = [g_N \, \phi(T;m_N)]^{-2} \, \frac{1}{2\pi^3} \, \int_{2m_N}^{\infty} d\varepsilon \, \varepsilon^2 \, \operatorname{K}_1(\varepsilon/T) \nonumber \\
& \qquad \times
\sum_{T=0,1} \, \sum_{S=0,1} \, \sum_{L} \, \sum_{J=|L-S|}^{L+S}\, (2\,T + 1) \, (2\,J+1) \, \delta_L^{\rm hc} (\varepsilon).
}

Let us denote the BU approach with hard-core interaction potential as BU-HC.
As follows from Eq.~\eqref{eq:b2ev}, the BU-HC approach  predicts a temperature dependent excluded volume parameter $v_{NN} (T) = -b_2^{NN} (T)$, at least on the level of the 2nd order virial expansion.

The coefficient $b_2^{NN}$ contains contributions from proton-proton, proton-neutron, and neutron-neutron scatterings.
It is also possible to calculate, separately, the 2nd cluster integral $b_2^{pp}$ for a pure proton system. It coincides with the $b_2^{nn}$ coefficient of the pure neutron system due to isospin symmetry.
The isospin quantum number is then not needed, and $b_2^{pp}$ reads
\eq{\label{eq:BUpp}
b_2^{pp}(T) = [(g_N/2) \, \phi(T;m_N)]^{-2} \, \frac{1}{2\pi^3} \, \int_{2\,m_N}^{\infty} d\varepsilon \, \varepsilon^2 \, \operatorname{K}_1(\varepsilon/T) \sum_{S=0,1} \, \sum_{L} \, \sum_{J=|L-S|}^{L+S}\, (2\,J+1) \, \delta_L^{\rm hc} (\varepsilon).
}

It is also useful to consider the original, non-relativistic BU formula~\cite{Beth:1937zz},
\eq{\label{eq:bNNnonrel}
b_2^{NN,\rm nr}(T) & = [g_N \, \phi^{\rm nr}(T;m_N)]^{-2} \, \frac{(2\,m_N)^2 \,}{2\pi^3} \, \sqrt{\frac{\pi T}{4 m_N}} \, \exp\left(-\frac{2\,m_N}{T}\right) \, \int_{0}^{\infty} d\varepsilon \, \exp(-\varepsilon/T) \nonumber \\
& \qquad \times
\sum_{T=0,1} \, \sum_{S=0,1} \, \sum_{L} \, \sum_{J=|L-S|}^{L+S}\, (2\,T + 1) \, (2\,J+1) \, \delta_L^{\rm hc} (\varepsilon),
}
where
\eq{
\phi^{\rm nr}(T;m) = \left( \frac{mT}{2\pi}\right)^{3/2} \, \exp\left( -\frac{m}{T} \right).
}
A comparison of the non-relativistic BU-HC result~\eqref{eq:bNNnonrel} with the classical result~\eqref{eq:evcl} 
provides an important cross check. For high temperatures the quantum effects in the BU-HC  model become unimportant,
thus, the results \eqref{eq:bNNnonrel} and~\eqref{eq:evcl} should coincide.

\subsection{Calculation results}

\begin{figure}[t]
\centering
\includegraphics[width=0.70\textwidth]{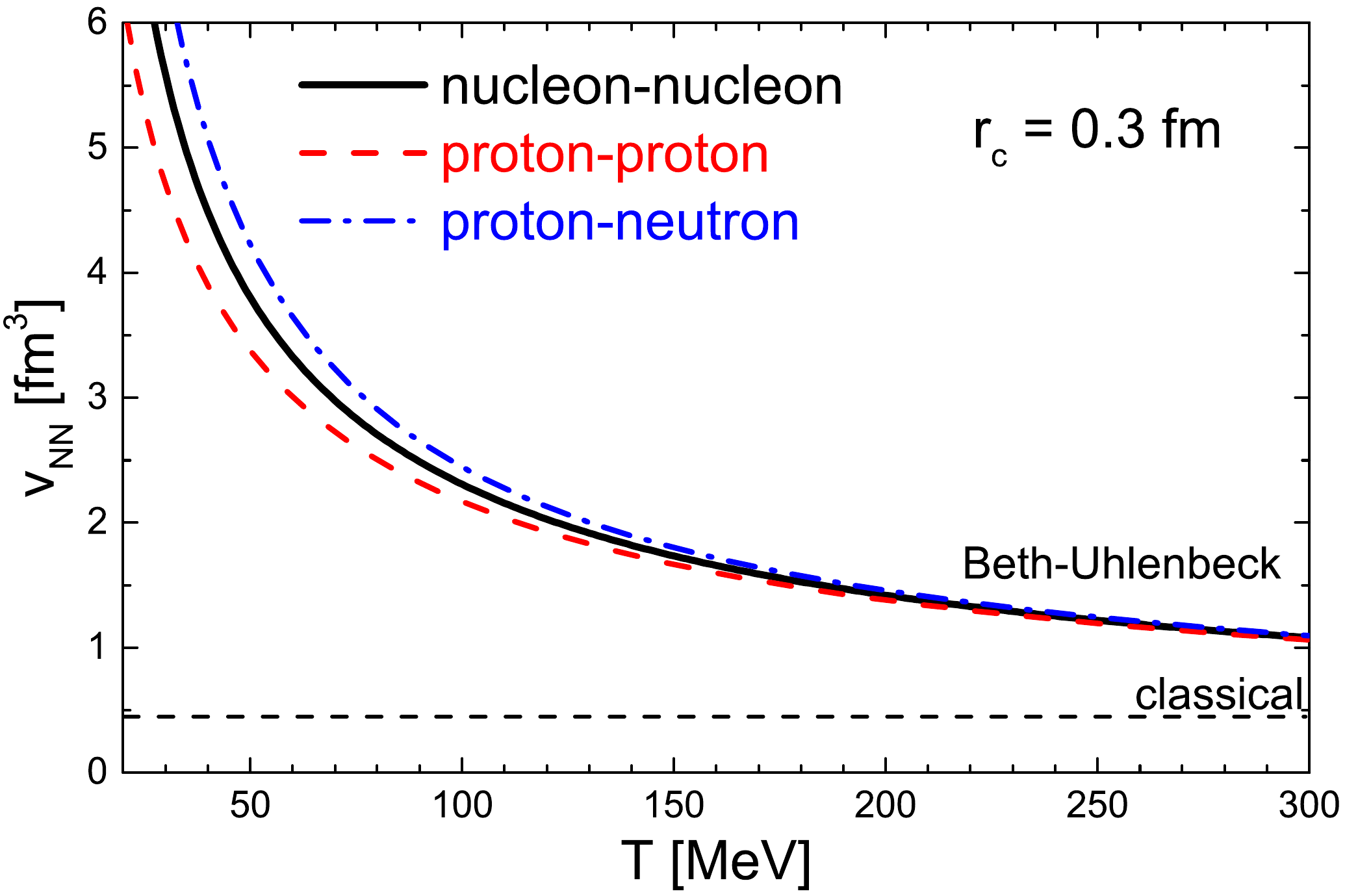}
\caption[]{
The temperature dependence of the nucleon-nucleon excluded volume parameter $v_{\rm NN}$ (solid black line), the proton-proton excluded volume parameter $v_{pp}$ (dashed red line), and the proton-neutron excluded volume parameter ${v_{pn} \equiv 2 \, v_{\rm NN} - v_{pp} }$ (dashed red line), as calculated within the relativistic Beth-Uhlenbeck approach for a hard-core potential with the nucleon hard-core radius of $r_c = 0.3$~fm.
The dashed horizontal line shows the prediction of the classical EV model \eqref{eq:evcl} with the same value of $r_c = 0.3$~fm.
}\label{fig:vNN}
\end{figure}

Figure~\ref{fig:vNN} depicts the temperature dependence of the nucleon-nucleon excluded volume parameter $v_{NN}$, calculated using Eq.~\eqref{eq:BUNN2} for the nucleon hard-core radius of $r_c = 0.3$~fm for temperatures up to $T = 300$~MeV. The temperature dependences of the proton-proton eigenvolume $v_{pp}$ and of the proton-neutron eigenvolume ${v_{pn} = 2 \, v_{NN} - v_{pp}}$ are depicted as well. The classical result~\eqref{eq:evcl} is depicted by the dashed horizontal line.
The numerical evaluation of Eq.~\eqref{eq:BUNN2} considers the terms with $L \leq 10$, and disregards the terms with $L > 10$. The higher order terms with $L > 10$ give a negligible contribution to $b_2^{\rm NN}$ for temperatures up to $T = 300$~MeV, as follows from numerical checks\footnote{In our figures
the results are presented up to rather high temperatures. This is done to see better a connection between different
model formulations. In reality, hadrons are not expected to be the dominant constituents of the strongly interacting matter at $T>200$~MeV.}.

Figure~\ref{fig:vNN} shows that the classical EV model~[Eq.~\eqref{eq:clev}] underestimates the value of the nucleon-nucleon excluded volume parameter by large factors of 3-4, at temperatures $T = 100-200$~MeV.
These temperature values are rather typical for the phenomenological applications of the EV model in the context of heavy-ion collisions and (Lattice) QCD equation of state.
Strong increase of $v_{NN}$ at low temperatures correlates with an increase of the thermal wavelength $\lambda_{dB}$. 

This result is quite remarkable:
the hard-core radii of hadrons are often used as an \emph{input} into the classical EV-HRG model, to describe repulsive interactions between hadrons at high densities~(see e.g. Refs.~\cite{Yen:1997rv,Yen:1998pa,BraunMunzinger:1999qy,Cleymans:2005xv,Andronic:2012ut,Begun:2012rf,Bhattacharyya:2013oya,Vovchenko:2014pka,Zalewski:2015yea,Vovchenko:2015cbk,Anchishkin:2014hfa,Kadam:2015xsa}).
A value $r_c = 0.3$~fm was sometimes taken based on the properties of nucleon-nucleon scattering~\cite{Andronic:2012ut,Bhattacharyya:2013oya}.
The large discrepancy between the classical EV model and the BU approach suggests that the former can only be considered as a simplified effective approach, when used in hadronic physics applications. This means that the parameter $a_2^{\rm ev}$ of the EV model should not be connected to the values of the hard-core radii via Eq.~(\ref{eq:evcl}). 
Note that similar concerns were voiced before, based on BU calculations for spinless particles~\cite{Kostyuk:2000nx,Typel:2016srf}.
More accurate analyses shall also take into account
interaction-channel dependent hard-core radii~\cite{Kostyuk:2000nx,Lo:2017ldt}.

\begin{figure}[t]
\centering
\includegraphics[width=0.70\textwidth]{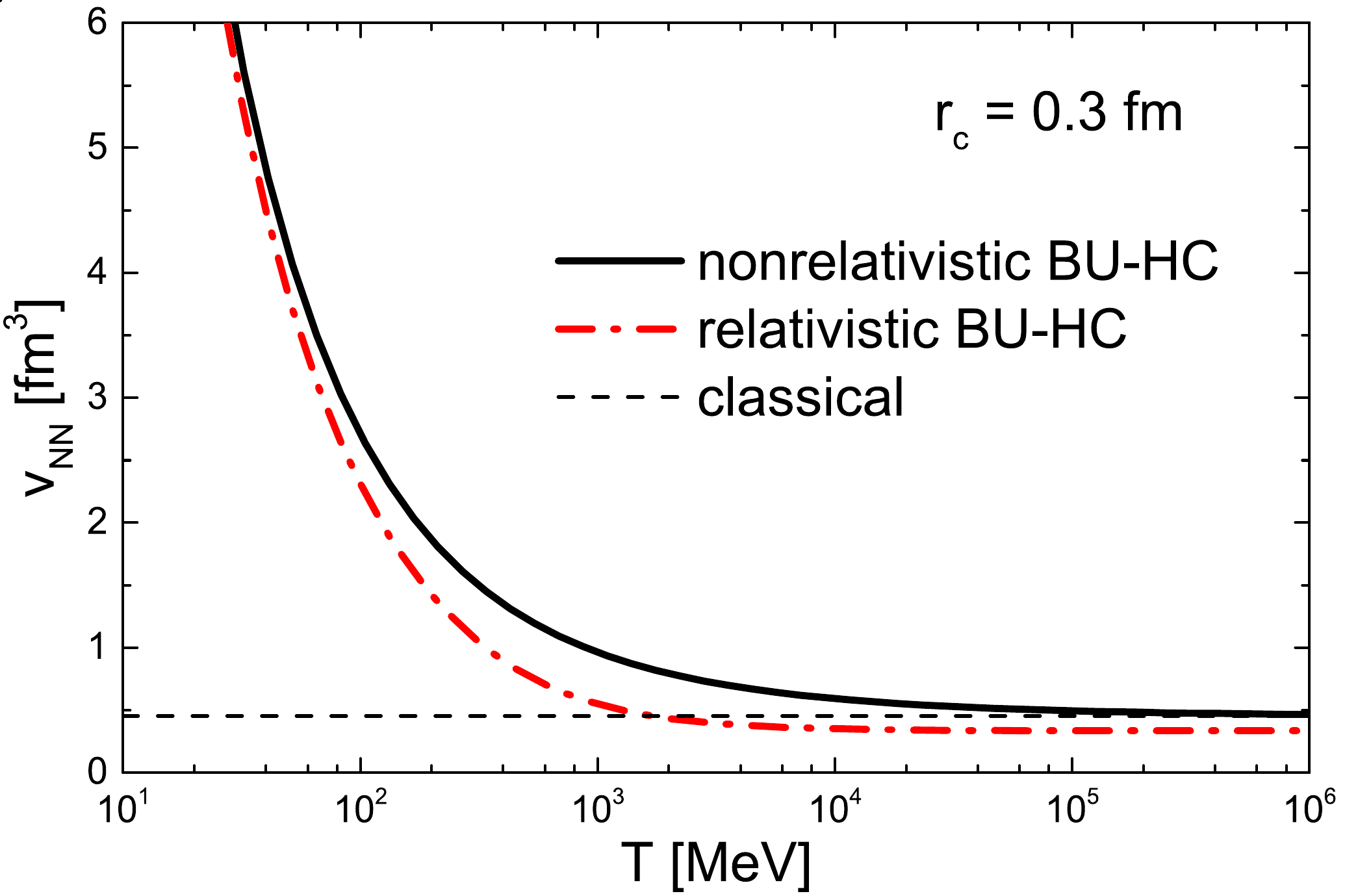}
\caption[]{
The temperature dependence of the nucleon-nucleon excluded volume parameter $v_{\rm NN}$ calculated using the non-relativistic~(solid black line) and relativistic~(dash-dotted red line) dispersion relations in the Beth-Uhlenbeck approach for hard core interaction potential, shown on the logarithmic temperature scale.
Nucleon hard-core radius of $r_c = 0.3$~fm is assumed.
The dashed horizontal line shows the prediction of the classical EV model~\eqref{eq:evcl} with the same value of $r_c = 0.3$~fm.
}\label{fig:vNNlogT}
\end{figure}

The classical EV model result [Eq.~\eqref{eq:clev}] is only valid when both, quantum mechanical and relativistic effects, can be neglected.
Formally, the non-relativistic BU-HC formula~\eqref{eq:bNNnonrel} is expected to converge to the classical result \eqref{eq:clev} at high temperatures.
This expectation was proven for spinless particles with a hard-core interaction ~\cite{Kostyuk:2000nx,Typel:2016srf}.
The numerical check for spin-$1/2$  nucleons is depicted in Figure~\ref{fig:vNNlogT}: The temperature dependence of the nucleon-nucleon excluded volume parameter $v_{NN}$, as calculated in the non-relativistic~(solid black line) and relativistic~(dash-dotted red line) BU-HC approach, for $r_c = 0.3$~fm, is shown on a logarithmic temperature scale, in the range $T = 10^1-10^6$~MeV.
Note that, in the present work, the difference between the relativistic and non-relativistic BU approaches is only in the dispersion relation between energy and momentum.
At very high temperatures, $T \sim 10^5$~MeV, the excluded volume parameter of the non-relativistic BU formula approaches the classical limit~(dashed line) from above, as expected.
These unrealistically high temperatures, however, are not relevant for any practical applications
since nucleons are expected to already melt into partons there.

The behavior of $v_{NN}(T)$ in the relativistic BU-HC approach~(dash-dotted red line) is similar to the non-relativistic BU-HC approach.
However, the relativistic approach yields systematically smaller values of $v_{NN}(T)$.
The limiting value of $v_{NN}(T)$ is slightly below the classical limit in the relativistic BU-HC approach.
Note that a relativistic formulation of the hard-core interaction problem
is not
fully consistent: The whole concept of a hard-core interaction is inconsistent with causality.
Nevertheless, nucleons are not affected that strongly by relativistic effects at temperatures which are discussed for the hadronic physics applications.
Therefore, the treatment of the hard-core repulsion between nucleons within the relativistic BU approach is considered satisfactory.

We note that scattering phase shifts can also be employed to study the non-equilibrium properties of interacting hadrons~\cite{Prakash:1993bt}. 
Therefore, one can study in a similar fashion the difference between classical and quantum mechanical hard-core repulsion for the various kinetic properties,
such as the scattering cross section and transport coefficients.
Similarly large differences could be expected there as well.
These extensions will be considered elsewhere.

\subsection{Other estimates and the role of attraction}

The results of the present approach can be compared to other estimates of the 2nd virial coefficient for nucleons.
These other estimates are not based on a hard-core interaction potential, at least not directly.
The 2nd virial coefficient should not be identified exclusively with an eigenvolume parameter in such a case, therefore we use the notation $a_2^{NN}$ instead of $v_{NN}$ for this comparison.
The comparison illustrates the relevance of the hard-core repulsion for thermodynamics of a nucleon gas.

\begin{figure}[t]
\centering
\includegraphics[width=0.70\textwidth]{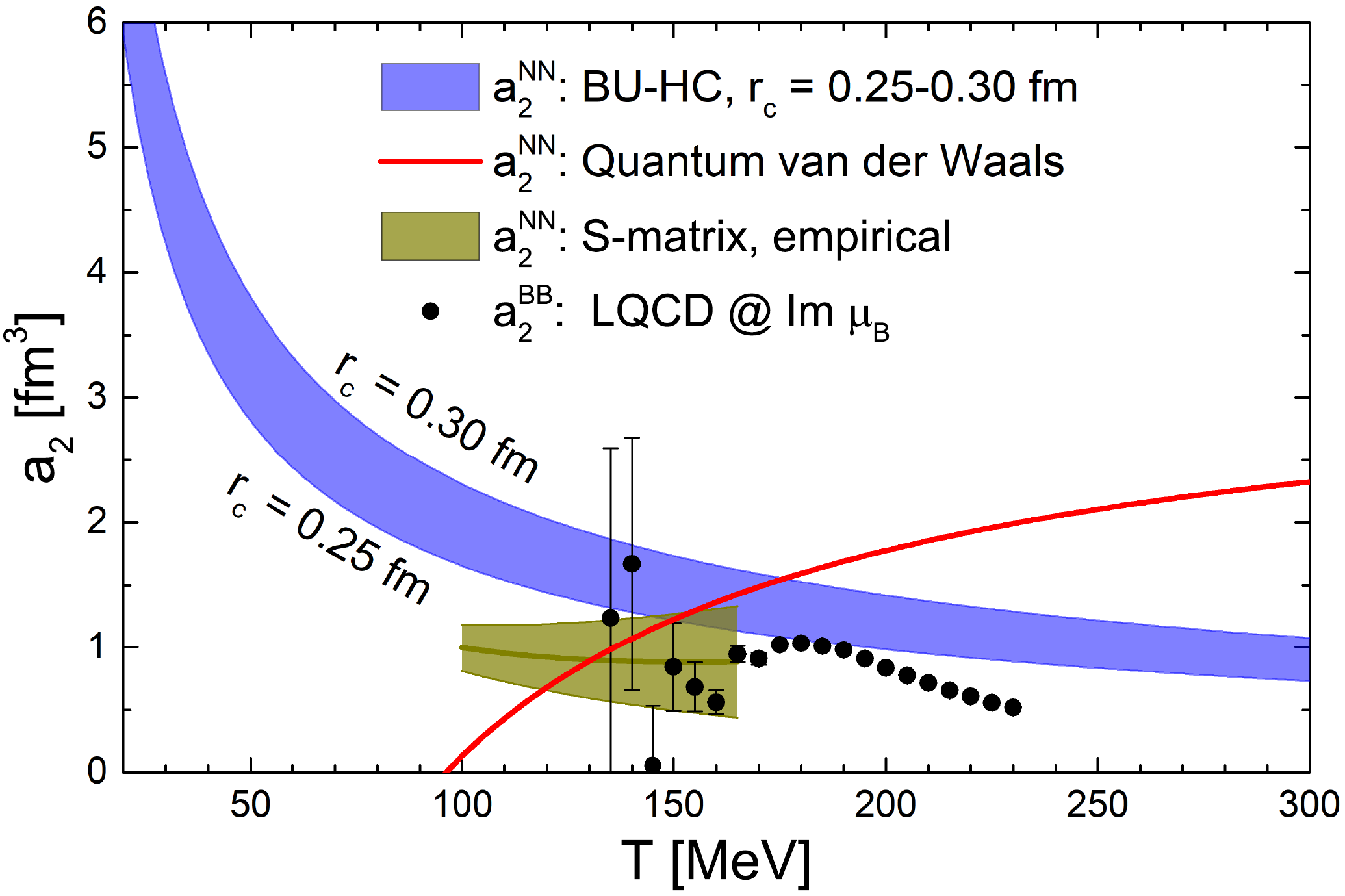}
\caption[]{
The temperature dependence of the second virial coefficient $a_2(T)$ of nucleon-nucleon interaction, calculated within different approaches.
The calculations within the relativistic Beth-Uhlenbeck approach for the system of nucleons with a hard-core interaction are depicted by the blue band, which results from the variation of the nucleon hard-core radius in the range $0.25 < r_c < 0.30$~fm.
The calculations of Ref.~\cite{Huovinen:2017ogf} within the S-matrix formalism, employing the empirical phase shifts of $NN$-scattering, are depicted by the yellow line with a band.
The red line depicts the second virial coefficient of nucleon-nucleon interaction in the quantum van der Waals model of nuclear matter~\cite{Vovchenko:2015vxa}.
Lattice QCD results for the 2nd virial coefficient of ``baryon-baryon interaction''~\cite{Vovchenko:2017xad}, obtained from simulations at an imaginary baryochemical potential, are depicted by black symbols with error bars.
}\label{fig:a2NN}
\end{figure}

For the hard-core repulsion, the empirical values of the nucleon hard-core radius $r_c$ are considered in the range $r_c = 0.25-0.30$~fm,
as suggested by the analysis of $NN$-scattering phase shift data~\cite{Wiringa:1994wb}.
The corresponding BU result is depicted in Fig.~\ref{fig:a2NN} by the blue band.
Decreasing $r_c$ from 0.3~fm to 0.25~fm results in about 30\% decrease of $a_2^{NN}(T)$ at a given temperature.

The present BU-HC approach accounts for the contribution of the short range repulsive hard-core interactions to the second virial  coefficient.
However, nucleon-nucleon interactions are also attractive at an intermediate range.
Attractive interactions give sizable negative contributions to $a_2^{NN}$. Especially at low temperatures, $T<20$~MeV, calculations~\cite{Horowitz:2005nd}, based on empirical phase shift data, do suggest that attractive interactions give the dominant contribution to $a_2^{NN}$.
Thus, the large positive contribution of the hard-core repulsion at low temperatures, as seen in Figs.~\ref{fig:vNN}-\ref{fig:a2NN}, is compensated by a similarly large, but negative contribution from the attraction.

A simple model which takes into account both attractive and repulsive interactions between nucleons is the Quantum van der Waals~(QvdW) model~\cite{Vovchenko:2015vxa}.
The QvdW model takes into account effects of Fermi statistics, important in the nuclear matter region at small temperatures and large baryon densities.
The repulsive and attractive interactions between nucleons are characterized by the temperature independent vdW parameters $b$ and $a$, respectively.
A fit to the nuclear ground state properties at $T=0$ yields values of $b = 3.42$~fm$^3$ and $a = 329$~MeV fm$^3$ for nucleons~\cite{Vovchenko:2015vxa}.
The second virial coefficient in this QvdW model reads\footnote{Once again, here we neglect the small ideal Fermi gas contribution to $a_2(T)$.} $a_2(T) = b - a/T$.
The temperature dependence of $a_2^{NN}(T)$ in the QvdW model is depicted in Fig.~\ref{fig:a2NN}, red line.
$a_2^{NN}$ is negative at small temperatures, crosses zero at $T = a / b \simeq 96$~MeV, and increases monotonically
at large temperatures.
This sign change of $a_2^{NN}(T)$ is expected for any system of interacting particles with short-range repulsion and intermediate range attraction.
At the same time, continued increase of $a_2^{NN}(T)$ at high temperatures in the QvdW model appears to be at odds with results of the BU-HC formalism. 
{This takes place 
because of the large, temperature independent value of the excluded-volume parameter $b$ in the QvdW model. Assuming $b=16\pi r_c^3/3$ one finds $r_c\cong 0.59$~fm.
This is essentially larger  than $r_c=0.2-0.3$~fm for the BU-HC results presented in Fig.~\ref{fig:a2NN}.

The second virial coefficient of the nucleon-nucleon interaction can be estimated in the S-matrix approach, by employing the empirically known phase shifts of $NN$-scattering.
This had recently been done in Ref.~\cite{Huovinen:2017ogf} for temperatures $100 < T < 165$~MeV.
The result is depicted by the yellow band in Fig.~\ref{fig:a2NN}.
The band itself results from the uncertainty in the contributions of the inelastic $NN$ channels to $a_2^{NN}(T)$.
The S-matrix result of Ref.~\cite{Huovinen:2017ogf} lies below our BU calculation, as expected, as the S-matrix calculation reflects the net contribution of attraction and repulsion between the nucleons to $a_2^{NN}(T)$.
The BU-HC calculation overestimates $a_2^{NN}$, as in the present work it manifests the repulsive hard-core interactions between nucleons only.
The difference between the present calculation and the S-matrix calculation of~Ref.~\cite{Huovinen:2017ogf} is reduced at higher temperatures: this reflects the fact that the short-range repulsive interactions dominate at higher temperatures.

For completeness, the recent imaginary-$\mu_B$ lattice QCD results on the partial pressure of QCD in the baryon number $|B| = 2$ sector~\cite{Vovchenko:2017xad} are also shown in Fig.~\ref{fig:a2NN} by black circles.
A purely hadronic description, reasonable at moderate temperatures, yields partial pressure proportional to an ``average'' second virial coefficient $a_2^{BB}$ for baryon-baryon interactions.
The error bars of the lattice estimations for $a_2^{NN}$ are rather large at $T < 160$~MeV.
The lattice results lie somewhat below the results of the BU calculations.

The comparisons shown in Fig.~\ref{fig:a2NN} suggest that the BU-HC calculation for $a_2^{NN}$ with $r_c = 0.25$~fm is quite consistent with  other estimates in the crossover temperature region, $T \sim 150$~MeV.
The BU-HC approach overestimates $a_2^{NN}$ at smaller temperatures due to the missing attractive interactions.
Therefore, modifications of the BU-HC approach are desirable for applications at these temperatures.

\section{Applications to the hadron resonance gas model}
\label{sec:HRG}

The BU-HC formalism is also useful to model the repulsive baryonic interactions in the 
HRG model.
Ref.~\cite{Vovchenko:2016rkn} considered an extension of the ideal HRG model where repulsive interactions act only between pairs of baryons and between pairs of antibaryons.
The system hence consists of three independent subsystems: non-interacting mesons, interacting baryons, and interacting antibaryons.
Thus, the pressure is given as the sum, $p = P_M + P_B + P_{\bar{B}}$.
It is assumed that the 2nd virial coefficient, $v_{BB} (T)$, which characterizes the baryon-baryon interactions, is the same for all (anti-)baryon pairs at a given temperature.
The nucleon-nucleon values, $v_{NN} (T)$, are taken for all
baryon-baryon and antibaryon-antibaryon pairs, i.e. $v_{BB} (T) \equiv v_{NN} (T)$.
This simplifying assumption is supported by lattice QCD simulations~\cite{Inoue:2011ai}, which do suggest that repulsive core is qualitatively similar between different baryon-baryon pairs.
The model probably overestimates the repulsive effects at high temperatures, where the high thermal pressure squeezes all hadron volumes~\cite{Kagiyama:1991uf,Ferroni:2008ej}.

The partial pressure of the baryonic and the antibaryonic subsystems in the BU-HC approach reads
\begin{subequations}
\label{eq:PBU}
\eq{\label{eq:PBBU}
P_B^{\rm BU} (T, \mu_B) & = T \, \phi_B(T) \, \lambda_B - T \, v_{BB} (T) \, [\phi_B(T) \lambda_B]^2, \\
\label{eq:PBBUbar}
P_{\bar{B}}^{\rm BU} (T, \mu_B) & = T \, \phi_B(T) \, \lambda_B^{-1} - T \, v_{BB} (T) \, [\phi_B(T) \lambda_B^{-1}]^2,
}
\end{subequations}
where $\lambda_B = \exp(\mu_B / T)$ and
\eq{
\phi_B (T) = \sum_{i \in B} \, \int d m \, \rho_i(m) \, \frac{d_i \, m^2 \, T}{2\pi^2} \, K_2\left( m \over T \right)
}
is the baryonic spectrum,
with $d_i$ and $\rho_i$ being, respectively, the degeneracy and a properly normalized mass distribution for hadron type $i$, and where the sum goes over all baryons in the system. We include all baryon states, which are listed as ``confirmed'' in the Particle Data Tables~\cite{Olive:2016xmw}.
The function $\rho_i$ takes into account the non-zero widths of the resonances
integrating over their Breit-Wigner shapes, following Refs.~\cite{Becattini:1995if,Wheaton:2004qb}.

The model given by Eq.~\eqref{eq:PBU} is dubbed BU-HRG,
the baryonic pressures \eqref{eq:PBBU} and \eqref{eq:PBBUbar}
contain only quadratic interaction terms, which are proportional to the 2nd cluster integral.
At large enough values of temperature and/or fugacity, the baryonic pressure will become negative, due to the negative sign of the quadratic term.
Thus, this pure BU approach is expected to break down at some point, namely when the higher order terms of the cluster expansion are no longer negligible.
It is instructive to consider the EV-HRG model with an effective temperature dependent excluded volume parameter.
The partial pressure of baryons and of antibaryons in such a model reads\footnote{The Fermi statistics effects are small in the considered temperature region and at $\mu_B=0$.}  
\begin{subequations}
\label{eq:Pev}
\eq{\label{eq:PBev}
P_B^{\rm ev}(T, \mu_B) & = T\, \phi_B(T) \, \lambda_B \, \exp\left(-~\frac{v_{BB} (T) \, P_B^{\rm ev}(T,\mu_B)}{T} \right), \\
\label{eq:PBevbar}
P_{\bar{B}}^{\rm ev}(T, \mu_B) & = T\, \phi_B(T) \, \lambda_B^{-1} \, \exp\left(-~\frac{v_{BB} (T) \, P_{\bar{B}}^{\rm ev}(T,\mu_B)}{T} \right).
}
\end{subequations}
It can be easily seen that the pressure~\eqref{eq:Pev} of the EV-HRG model is consistent with the BU approach~\eqref{eq:PBU} up to the second order of the cluster expansion.
However, the EV-HRG model also contains non-zero higher order coefficients in the cluster expansion.
Hence, large differences between the two models
may indicate that
the second order cluster expansion
is not applicable any longer.

Consider the temperature dependence of the baryon susceptibilities at $\mu_B = 0$: the $n$-th order baryon susceptibility $\chi_n^B$ is defined as
\eq{
\chi_n^B = \left.\frac{\partial^n (p/T^4)}{\partial (\mu_B/T)^n}\right|_{\mu_B = 0}~.
}

These
higher-order susceptibilities are a sensitive measure of
the response of the system to changes in the $\mu_B/T$ values, and are especially sensitive to the various baryon-baryon interactions.
Consider the effects of the repulsive hard-core interactions between baryons on these observables:
The BU-HC calculations of $v_{NN}(T)$ for nucleons with $r_c = 0.25-0.3$~fm, presented in the previous section and depicted by the blue band in Fig.~\ref{fig:a2NN}, are
used for $v_{BB}(T)$ in Eqs.~(\ref{eq:PBev}) and (\ref{eq:PBevbar}).

\begin{figure}[t]
\begin{center}
\includegraphics[width=0.49\textwidth]{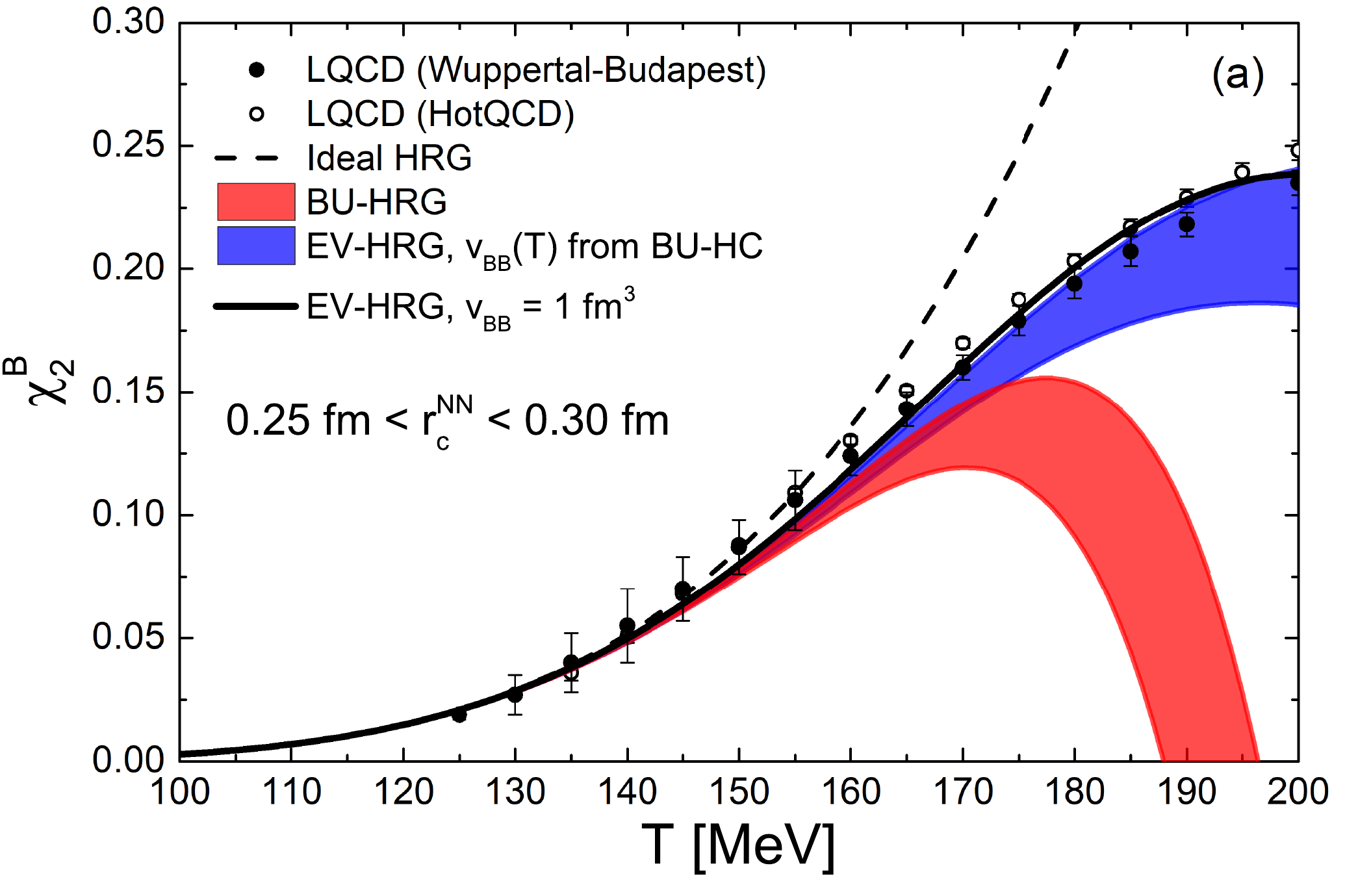}
\includegraphics[width=0.49\textwidth]{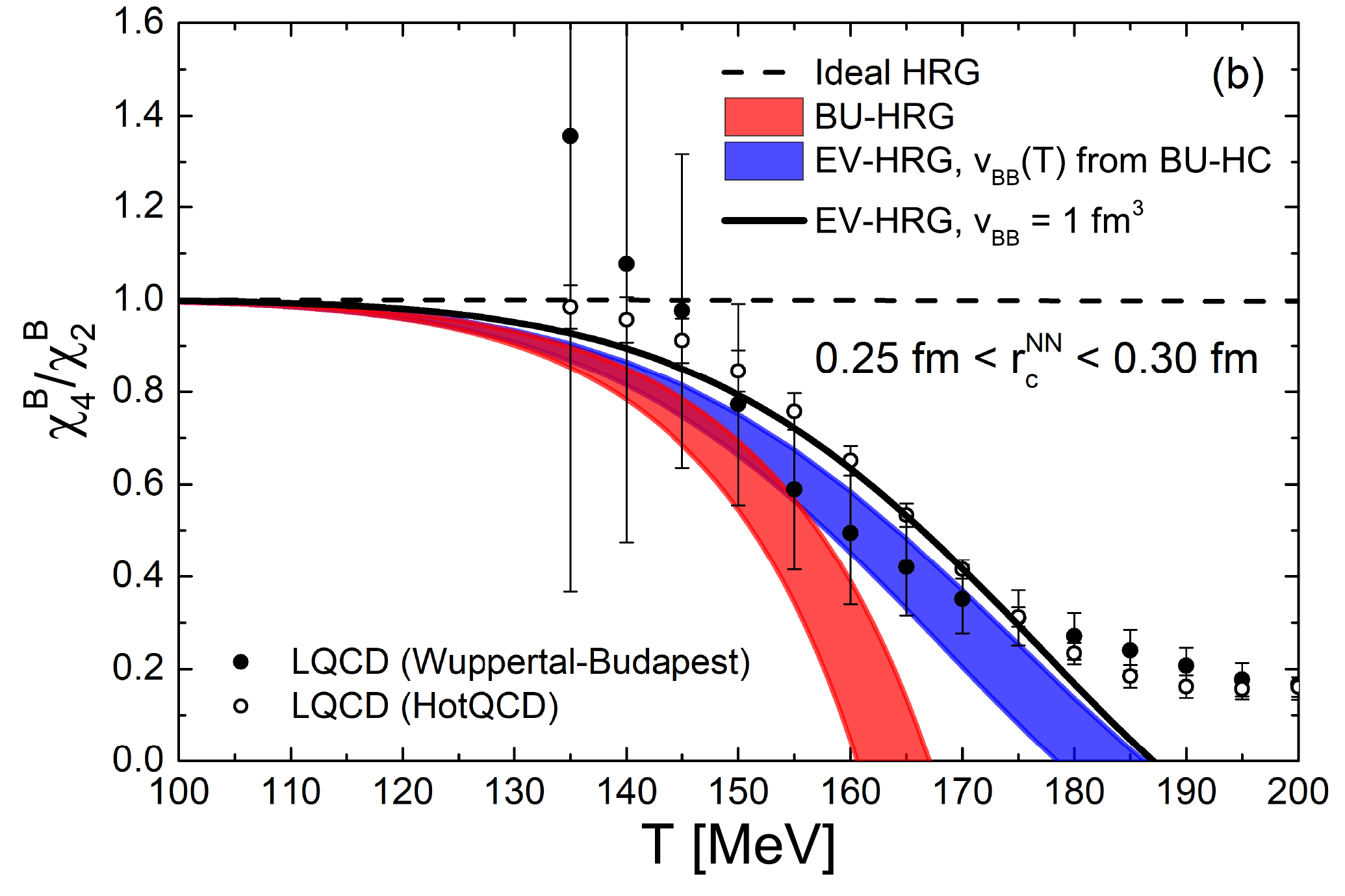}
\caption{The temperature dependence of
(a) $\chi_2^B$ and (b) $\chi_4^B / \chi_2^B$ net baryon number susceptibilities at $\mu_B = 0$, as calculated within the I-HRG model (dashed black lines), the BU-HRG model (red bands), and the EV-HRG model (blue bands) with the temperature dependent baryon excluded volume parameter, using for all (anti)baryons the Beth-Uhlenbeck value for nucleons.
The bands result from the variation of the nucleon hard-core radius in the range $r_c \simeq 0.25-0.3$~fm.
The lattice QCD results of the Wuppertal-Budapest~\cite{Borsanyi:2011sw,Bellwied:2015lba} and HotQCD~\cite{Bazavov:2017dus,Bazavov:2017tot} collaborations
are shown by the full and open symbols, respectively. Solid lines correspond to the EV-HRG model with $v_{BB}=1$~fm$^3$.
}
\label{fig:chiB}
\end{center}
\end{figure}

The resulting $\chi_2^B (T)$ and $\chi_4^B (T) / \chi_2^B (T)$ are depicted in Fig.~\ref{fig:chiB}. The red bands correspond to the BU-HRG model
(\ref{eq:PBBU},\ref{eq:PBBUbar}), the blue bands depict the EV-HRG model (\ref{eq:PBev},\ref{eq:PBevbar}), and the ideal HRG model results are shown by the dashed lines.
The lattice QCD results of the Wuppertal-Budapest~\cite{Borsanyi:2011sw,Bellwied:2015lba} and HotQCD~\cite{Bazavov:2017dus,Bazavov:2017tot} collaborations
are shown by the full and open symbols, respectively.
At low temperatures, $T \lesssim 110$~MeV, the effect of the repulsive interactions on $\chi_2^B (T)$ and $\chi_4^B (T) / \chi_2^B (T)$ is negligible.
This is in spite of the strong increase of the excluded-volume parameter in the BU-HC approach at low temperatures.
The effect is small because of an exponential decrease of the density of baryons, which renders the influence of baryonic interactions negligible at low temperatures and $\mu_B = 0$.
Repulsive baryon-baryon interactions suppress baryon susceptibilities at higher temperatures, as compared to the ideal HRG result.
At moderate temperatures, $T \lesssim 150$~MeV for $\chi_2^B$, and $T \lesssim 130$~MeV for $\chi_4^B/\chi_2^B$, this suppression is described nearly identically in BU-HRG and EV-HRG models.
The total densities of baryons and of antibaryons at $\mu_B = 0$ increase strongly as the temperature is increased.
Higher terms of the cluster expansion are therefore non-negligible at higher temperatures.
This is reflected in larger differences between the predictions of the BU-HRG and the EV-HRG models at $T \gtrsim 160$~MeV.
It is seen from Fig.~\ref{fig:chiB}a that $\chi_2^B$ is negative at $T \gtrsim 190$~MeV in the BU-HRG model.
By definition, $\chi_2^B$ characterizes the width of the fluctuations of the net baryon number.
The negative values of $\chi_2^B$ in the BU-HRG model are unphysical -- they simply characterize the breakdown of the second order virial expansion at high temperatures.

In contrast, the EV-HRG model predicts a reasonable behavior of the baryon number susceptibilities even at high temperatures.
The EV-HRG calculations with $v_{BB}(T)=v_{NN}(T)$ calculated within BU-HC approach for $r_c = 0.25$~fm give an overall satisfactory description of the lattice data up to $T \simeq 175-180$~MeV.
The deviations of the ideal HRG model from lattice QCD data for the baryon susceptibilities
in the vicinity and even somewhat above the pseudocritical temperature can be understood in terms of the repulsive baryonic interactions.
This conclusion was reported previously in Refs.~\cite{Vovchenko:2016rkn,Vovchenko:2017cbu,Huovinen:2017ogf}.

The underestimation of the lattice data at $T \sim 150-160$~MeV is attributed to the missing attractive interactions between baryons in the BU-HC calculation of $v_{NN}(T)$, as discussed in the previous section.
A possible way to take into account the residual attraction between baryons is to rescale $v_{NN}(T)$ to smaller values, and then use these values in the EV-HRG model.
Calculations of $\chi_2^B$ and $\chi_4^B / \chi_2^B$ within the EV-HRG model with a constant temperature independent value $v_{NN} = 1$~fm$^3$,
motivated by the $a_2^{NN}$ estimates
in Fig.~\ref{fig:a2NN},
are
depicted in Fig.~\ref{fig:chiB}~(solid lines). 
This further improvement of the description of the lattice data in the crossover region
by the EV-HRG model with 
$v_{NN} = 1$~fm$^3$ provides effectively a good
approximation of the quantum
description of
baryon-baryon interactions in the crossover temperature region.
Thus, this model can be used for interpretation of the lattice QCD data; the model is also quite reasonable for the thermal analysis of baryon-related observables in heavy-ion collision experiments.
Note that the value $v_{NN} = 1$~fm$^3$ was also suggested in the recent analysis of the lattice QCD data at imaginary baryochemical potential~\cite{Vovchenko:2017xad}.

\section{Summary}
\label{sec:summary}

The quantum mechanical Beth-Uhlenbeck treatment of the hard-core interactions between nucleons/baryons presented here has proven to be a clear progress as compared to the simple classical approach,
as it remedies many of the formerly ununderstood discrepancies between lattice QCD calculations and the common ideal hadron resonance gas model.

The Beth-Uhlenbeck approach yields a strongly temperature dependent second virial coefficient of nucleon-nucleon interactions, which can be interpreted as a temperature dependent excluded volume parameter.
The classical EV model underestimates the value of the nucleon-nucleon excluded volume parameter by factor 3-4 at temperatures $T = 100-200$~MeV for a given value of the nucleon hard-core radius $r_c$.
Such temperature range values are typical in in the EV model applications for fitting the heavy-ion collision data and studying the QCD equation of state.
These large discrepancies suggest that the classical EV model is only an effective approach -- when used in hadronic physics applications,
the effective radius parameters are strongly modified.
Attempts to connect the values of the 2nd virial coefficients of various hadron-hadron interactions in any EV-type approach, to the corresponding hard-core radii of hadrons must consider these discrepancies.
In particular, those EV-models which fix the radii parameters
on the basis of the empirical knowledge of the hard-core radius of nucleon-nucleon interaction~\cite{BraunMunzinger:1999qy,Andronic:2012ut,Bhattacharyya:2013oya}
should be re-evaluated.

The temperature dependent excluded volume parameter for nucleons is calculated in the Beth-Uhlenbeck approach,
assuming hard-core radii of $r_c = 0.25-0.3$~fm, as suggested by the empirical data.
This parameter range is then used to model the repulsive baryonic interaction in the hadron resonance gas model. The predictions
for net baryon number susceptibilities are compared to the lattice QCD calculations.
It is found that this modified Beth-Uhlenbeck approach describes fairly well the
deviations of the lattice data from the ideal HRG model
at $T \lesssim 160$~MeV.
The model breaks down at higher temperatures due to the absence of the non-negligible higher-order terms of the cluster expansion.
The excluded volume HRG model with the temperature dependent baryonic eigenvolume, on the other hand, extends the agreement with the lattice data for baryon number susceptibilities even to the temperatures beyond 160~MeV.

Finally, one should note that the intermediate range attractive baryonic interactions, neglected in the present Beth-Uhlenbeck calculations, influence the thermodynamics of a hadron gas.
Effects of attractive interactions are strong at low temperatures and residual at high temperatures.
Present analysis implies, that the excluded volume HRG model with a constant effective baryonic ``excluded-volume'' parameter $v_{BB} = 1$~fm$^3$
provides a simple yet efficient 
description of the net effect of the repulsive and attractive baryon-baryon interactions on the hadronic equation of state in the crossover temperature region.

\vspace{0.5cm}
\section*{Acknowledgements}
We are grateful to J.~Steinheimer, P.~Alba, K.~Redlich, J.~Cleymans, B.~Friman, and in particular to P.M.~Lo, for useful discussions regarding the differences between the excluded volume and Beth-Uhlenbeck approaches to treat repulsive hadronic interactions.
We thank P.~Petreczky for providing the data on the $S$-matrix calculated $NN$ virial coefficient in a tabulated format, and for useful comments.
We also thank F.~Karsch and S.~Mukherjee for providing the lattice data of the HotQCD collaboration.
This work was supported by the Helmholtz International Center for FAIR within the LOEWE program of the State of Hesse.
V.V. and A.M. acknowledge the support by HGS-HIRe for FAIR.
A.M. is furthermore thankful for the
support by the Norwegian Centre for International
Cooperation in Education (SIU) for financial support, grant ``CPEA-LT-2016/10094 From
Strong Interacting Matter to Dark Matter''.
The work of M.I.G. is supported 
by the Program of Fundamental Research of the Department of Physics
and Astronomy of National Academy of Sciences of Ukraine.
H.St. appreciates the J.M.~Eisenberg Laureatus endowed chair of the Fachbereich Physik at Goethe University, and the Walter Greiner Gesellschaft, Frankfurt.

\end{document}